\documentclass[10pt, conference, twocolumn]{IEEEtran}
\usepackage{times}
\usepackage{graphicx}
\usepackage{hyperref}
\usepackage{cite}
\usepackage{amsmath}
\usepackage{booktabs}
\usepackage{balance} 

\title{CveBinarySheet: A Comprehensive Pre-built Binaries Database for IoT Vulnerability Analysis}

\author{
    \IEEEauthorblockN{Lingfeng Chen}
    \IEEEauthorblockA{Independent Researcher\\
                      Email: \href{mailto:lingfengchen@webster.edu}{lingfengchen@webster.edu}}
}

\begin{document}

\maketitle

\begin{abstract}
    Binary Static Code Analysis (BSCA) is a pivotal area in software vulnerability research, focusing on the precise localization of vulnerabilities within binary executables. Despite advancements in BSCA techniques, there is a notable scarcity of comprehensive and readily usable vulnerability datasets tailored for diverse environments such as IoT, UEFI, and MCU firmware. To address this gap, we present CveBinarySheet, a meticulously curated database containing 1033 CVE entries spanning from 1999 to 2024. Our dataset encompasses 16 essential third-party components, including \texttt{busybox} and \texttt{curl}, and supports five CPU architectures: x86-64, i386, MIPS, ARMv7, and RISC-V64. Each precompiled binary is available at two compiler optimization levels (O0 and O3), facilitating comprehensive vulnerability analysis under different compilation scenarios. By providing detailed metadata and diverse binary samples, CveBinarySheet aims to accelerate the development of state-of-the-art BSCA tools, binary similarity analysis, and vulnerability matching applications.
\end{abstract}


\section{Introduction}
\label{sec:introduction}

The proliferation of Internet of Things (IoT) devices has introduced significant security challenges, primarily due to the diverse and often constrained environments in which these devices operate. Firmware vulnerabilities can lead to severe security breaches, making the identification and analysis of such vulnerabilities paramount. Binary Static Code Analysis (BSCA) is a pivotal area of research within the domain of software vulnerability analysis. As the demand for higher precision in vulnerability localization intensifies, current methodologies for identifying vulnerabilities within binary executable files predominantly operate at the function level. This granularity allows for more accurate pinpointing of vulnerable code segments, facilitating targeted mitigation strategies.

However, despite the advancements in BSCA techniques, there remains a significant gap in the availability of comprehensive and readily usable vulnerability datasets tailored for binary analysis. Existing research efforts often lack accessible datasets that encompass a wide range of vulnerabilities across multiple CPU architectures, particularly those pertinent to IoT, UEFI, and MCU firmware environments. The scarcity of such datasets hampers the development and benchmarking of state-of-the-art (SOTA) binary analysis tools, limiting their effectiveness in real-world applications.

To bridge this gap, we introduce CveBinarySheet, a meticulously curated database of pre-built binaries associated with 1033 CVE entries, spanning multiple CPU architectures commonly found in IoT, UEFI, and MCU firmware. Our dataset builds upon existing resources like MegaVul \cite{ni2023megavul}, advancing the field by focusing specifically on user-space third-party components and providing compiled binary executables. This ensures comprehensive coverage of commonly used software dependencies that may harbor vulnerabilities and facilitates more accurate and efficient vulnerability analysis.

This paper details the construction of CveBinarySheet, outlining the methodologies employed to compile vulnerable binaries and extract detailed vulnerability metadata. By providing a comprehensive and diverse set of pre-built binaries, CveBinarySheet aims to empower researchers and developers to advance BSCA, binary similarity analysis, and vulnerability matching tools. 

\section{Construction}
\label{sec:construction}

\subsection{CVE Coverage}
CveBinarySheet encompasses 1033 CVE entries, spanning vulnerabilities disclosed from 1999 to 2024. Each CVE is meticulously documented with comprehensive metadata to facilitate in-depth vulnerability analysis:
\begin{itemize}
    \item \textbf{CVE Identifier}: Unique identifier (e.g., CVE-2021-XXXX).
    \item \textbf{Patch URL}: Links to patches or commits addressing the vulnerability.
    \item \textbf{Reaching Path}: URLs to additional information, typically from the National Vulnerability Database (NVD).
    \item \textbf{Function Names}: Lists of functions affected by the vulnerability, extracted using Tree-sitter.
    \item \textbf{Affected Versions}: Specifies the range of software versions impacted.
    \item \textbf{Binary Version}: Indicates the version used to compile the provided binary.
    \item \textbf{File Changes}: Details of file paths and line numbers altered before and after patching, extracted via Tree-sitter.
\end{itemize}

\subsection{Supported Architectures}
CveBinarySheet includes pre-built binaries for the following five CPU architectures, covering a broad spectrum of devices and firmware types:
\begin{itemize}
    \item x86-64
    \item i386
    \item MIPS
    \item ARMv7
    \item RISC-V64
\end{itemize}
These architectures are prevalent in IoT devices, UEFI firmware, and MCU firmware, ensuring wide applicability for security analysis. Notably, RISC-V64 has been specifically considered due to its growing adoption in MCU and other embedded domains. This consideration ensures that the dataset accommodates the unique binary executable formats associated with RISC-V architectures, thereby enhancing its relevance for emerging firmware environments.

\subsection{Third-Party Components}
CveBinarySheet incorporates pre-built versions of 16 essential third-party components, enhancing the dataset's robustness and applicability. These components include, but are not limited to:
\begin{itemize}
    \item \texttt{busybox}
    \item \texttt{coreutils}
    \item \texttt{curl}
    \item \texttt{ffmpeg}
    \item \texttt{imagemagick}
    \item \texttt{libpcap}
    \item \texttt{libpng}
    \item \texttt{libtiff}
    \item \texttt{libuv}
    \item \texttt{libxml2}
    \item \texttt{lua}
    \item \texttt{mbedtls}
    \item \texttt{mutt}
    \item \texttt{openvpn}
    \item \texttt{openssl}
    \item \texttt{wget}
\end{itemize}
These components are widely used in IoT and firmware environments, making them relevant targets for vulnerability analysis. By including these third-party libraries and tools, CveBinarySheet ensures comprehensive coverage of common software dependencies that may harbor vulnerabilities.

\subsection{Choice of Compilation Environment}
To ensure consistency and reliability in the compiled binaries, we utilize Arch Linux's Arch User Repository (AUR) for the compilation process. The decision to adopt AUR is driven by several key factors:

\begin{itemize}
    \item \textbf{Simplicity and Ease of Use}: AUR provides a vast collection of user-contributed packages, simplifying the process of accessing and compiling a wide range of software components. Its straightforward package management system allows for efficient installation and maintenance of necessary dependencies.
    
    \item \textbf{Rolling Release Model}: Arch Linux follows a rolling release paradigm, ensuring that packages are continuously updated with the latest features and security patches. This approach minimizes the likelihood of encountering outdated dependencies or vulnerabilities that may arise from fixed release cycles found in other distributions like Debian.
    
    \item \textbf{Reduced Upstream Patch Interference}: The rolling update mechanism of Arch Linux ensures that the system remains up-to-date without requiring major version upgrades. Consequently, the compiled binaries are less susceptible to disruptions caused by upstream patches, maintaining their reproducibility and stability over time.
    
    \item \textbf{Flexibility for Custom Configurations}: AUR's user-driven nature allows for greater flexibility in configuring compilation settings, enabling the creation of binaries tailored to specific research requirements. This flexibility is essential for accurately reproducing vulnerable states across different CPU architectures and optimization levels.
\end{itemize}

In contrast, distributions like Debian adhere to more static release cycles, which can introduce dependencies and patches that may interfere with the reproducibility of vulnerable binaries. By leveraging Arch Linux's AUR, we achieve a balance between ease of package management and the stability required for compiling consistent and reliable binary executables.

\section{Data Category}
\label{sec:data_category}

\subsection{Classification of Pre-built Binaries}
CveBinarySheet organizes its pre-built binary executables in a hierarchical structure based on several key attributes: component name, version number, CPU architecture, and compiler optimization level. This classification scheme ensures that binaries are systematically categorized, facilitating easy retrieval and analysis. The hierarchical categorization is structured as follows:

\begin{itemize}
    \item \textbf{Component Name}: Binaries are first grouped by the third-party component they represent, such as \texttt{busybox}, \texttt{curl}, etc. This grouping allows researchers to focus on specific components of interest within the dataset.
    \item \textbf{Version Number}: Within each component group, binaries are further categorized by their specific version numbers. This granularity is crucial for analyzing how vulnerabilities evolve across different software versions.
    \item \textbf{CPU Architecture}: Each version of a component is compiled for multiple CPU architectures, including x86-64, i386, MIPS, ARMv7, and RISC-V64. This multi-architecture support ensures that the dataset caters to a wide range of hardware platforms prevalent in IoT, UEFI, and MCU firmware environments.
    \item \textbf{Compiler Optimization Level}: For each binary, two distinct compiler optimization levels are provided: O0 (no optimization) and O3 (high optimization). Offering binaries at these two optimization levels allows researchers to study the impact of compiler optimizations on vulnerability detection and binary analysis techniques.
\end{itemize}

This hierarchical classification not only streamlines the dataset organization but also enhances its utility for diverse research applications, enabling targeted analyses based on specific components, versions, architectures, or optimization settings.

\subsection{Compilation Scripts Classification}
The compilation scripts in CveBinarySheet are meticulously organized to support the reproducibility and scalability of binary generation. These scripts are classified based on component name and version number, following a hierarchical structure that mirrors the binary classification. The classification details include:

\begin{itemize}
    \item \textbf{Component Name}: Similar to the binary classification, compilation scripts are first grouped by the third-party component they correspond to. This ensures that all scripts related to a specific component are easily accessible.
    \item \textbf{Version Number}: Within each component group, scripts are further organized by the version numbers of the component. This organization allows for precise reproduction of binaries corresponding to specific software versions.
\end{itemize}

Each compilation script is designed to automate the build process for a given component and version, handling dependencies and configuration settings required for successful compilation across different CPU architectures and optimization levels. 

\subsection{Integration with Existing Categories}
The classification schemes for pre-built binaries, CVE information, and compilation scripts are interlinked, providing a cohesive structure that enhances the dataset's overall organization. This integration ensures that users can navigate the dataset intuitively, accessing binaries, scripts, and vulnerability details in a synchronized manner. By maintaining consistent categorization across different data facets, CveBinarySheet promotes a unified and efficient research workflow, accommodating a wide range of analysis and evaluation tasks.

\section{Further Application}
\label{sec:further_application}

CveBinarySheet serves as a foundational resource for various security research applications. Below, we outline three primary application directions where this dataset can significantly contribute:

\begin{enumerate}
    \item \textbf{Training Vulnerability Similarity Matching Models}:
    \begin{itemize}
        \item Recent advancements in Binary Static Code Analysis (BSCA) have leveraged sophisticated machine learning models such as BERT and Transformers to enhance binary function similarity detection. For instance, \textbf{jTrans} \cite{wang2022jtrans} employs a Transformer-based architecture to learn representations of binary code, incorporating control flow information to improve similarity detection accuracy.
        \item By employing CveBinarySheet, researchers can expedite the development of similarity analysis models. The comprehensive collection of 1033 CVEs across multiple architectures provides a robust dataset for training and evaluating models in realistic scenarios.
        \item This dataset enables the rapid assessment of model performance in detecting vulnerabilities across diverse firmware environments, enhancing the applicability and reliability of similarity matching tools in real-world applications.
    \end{itemize}
    
    \item \textbf{Training Vulnerability Repair Corpora based on LLM}:
    \begin{itemize}
        \item CveBinarySheet includes detailed JSON metadata that outlines code modifications before and after vulnerability patches, including specific line numbers. This information is crucial for training models to understand and predict vulnerability fix patterns.
        \item Studies such as \textbf{"Are Large Language Models Memorizing Bug Benchmarks?"} \cite{ramos2024large} have highlighted the importance of comprehensive patch data for training language models effectively. The availability of patch URLs and detailed code changes in our dataset facilitates the creation of high-quality training corpora for large language models.
        \item Leveraging this data, researchers can train models to generate accurate and context-aware code fixes, improving automated vulnerability remediation tools.
    \end{itemize}
    
    \item \textbf{Application in Binary-Level Formal Analysis}:
    \begin{itemize}
        \item Tools like \textbf{CWE Checker} \cite{barabosch2023cwechecker} perform formal analysis of binary executables to identify vulnerabilities such as buffer overflows and use-after-free (UAF) errors. CveBinarySheet provides a diverse set of binaries that can be used to validate and benchmark such formal analysis tools.
        \item By utilizing our dataset, researchers can assess the effectiveness of formal analysis algorithms in detecting real-world vulnerabilities across different architectures and optimization levels.
        \item Additionally, CveBinarySheet can serve as a benchmark for developing new formal verification techniques, ensuring they are robust and generalizable across various firmware environments.
    \end{itemize}
\end{enumerate}

\section{Limitations}
\label{sec:limitations}

While CveBinarySheet offers a comprehensive collection of pre-built binaries across multiple architectures and compiler optimization levels, it is not without limitations:
\begin{itemize}
    \item \textbf{Coverage Gap}: The dataset includes 1033 CVEs, which, although substantial, may not cover all existing vulnerabilities, especially newly discovered ones post-2024.
    \item \textbf{Component Scope}: Only 16 third-party components are included, potentially excluding other widely used libraries and tools that may harbor vulnerabilities.
    \item \textbf{Compilation Consistency}: Variations in compilation environments beyond the specified optimization levels (e.g., different compiler versions or settings) might affect binary behavior and analysis results.
\end{itemize}

Future work aims to address these limitations by expanding the CVE coverage, incorporating additional third-party components, standardizing compilation environments further, and enhancing metadata extraction techniques.

\section{Related Work}
\label{sec:related_work}

\subsection{Application of BERT and Transformer in BSCA}
Recent advancements in Binary Static Code Analysis (BSCA) have leveraged sophisticated machine learning models such as BERT and Transformers to enhance binary function similarity detection. For instance, \textbf{jTrans} \cite{wang2022jtrans} employs a Transformer-based architecture to learn representations of binary code, incorporating control flow information to improve similarity detection accuracy. These models have demonstrated significant improvements over traditional approaches, achieving state-of-the-art performance in binary code similarity tasks. By utilizing contextual embeddings derived from these deep learning models, BSCA techniques can better capture the semantic and structural nuances of binary functions, facilitating more precise vulnerability localization and comparison across diverse architectures.

\subsection{Training Vulnerability Repair Corpora based on LLM}
Large Language Models (LLMs) have revolutionized various software engineering tasks, including automated vulnerability repair. By leveraging detailed patch information and code modifications provided in datasets like CveBinarySheet, researchers can train LLMs to generate accurate and context-aware code fixes. Studies such as \textbf{"Are Large Language Models Memorizing Bug Benchmarks?"} \cite{ramos2024large} have utilized comprehensive patch data to enhance the training of LLMs, enabling them to suggest effective patches for identified vulnerabilities. The availability of pre-built binaries along with precise vulnerability metadata in CveBinarySheet facilitates the creation of high-quality training corpora, thereby improving the performance and reliability of automated vulnerability repair systems.

\subsection{Application of LLM in BSCA}
LLMs have also found applications in BSCA beyond vulnerability repair, including automated code analysis and reverse engineering. These models can assist in generating detailed reports on binary executables, identifying potential security flaws, and suggesting mitigation strategies. By integrating LLMs with datasets like CveBinarySheet, researchers can develop more intelligent and autonomous BSCA tools capable of understanding complex binary structures and providing actionable insights. The combination of LLMs' natural language understanding and the structured vulnerability data in CveBinarySheet enhances the capability of BSCA systems to perform nuanced analyses, making them more effective in real-world security assessments.

\section{Conclusion}
\label{sec:conclusion}

CveBinarySheet addresses a critical need in the security research community by providing a comprehensive and diverse set of pre-built binaries associated with a wide range of CVEs across multiple architectures. By facilitating advanced binary analysis techniques, this dataset empowers researchers and developers to enhance the security of IoT devices, UEFI firmware, and MCU firmware. Future work includes expanding the dataset to encompass additional architectures and CVEs, as well as integrating automated analysis tools to further streamline vulnerability research.

\bibliographystyle{IEEEtran}
\bibliography{CveBinarySheet}

\begin{thebibliography}{1}
\providecommand{\url}[1]{#1}
\csname url@samestyle\endcsname
\providecommand{\newblock}{\relax}
\providecommand{\bibinfo}[2]{#2}
\providecommand{\BIBentrySTDinterwordspacing}{\spaceskip=0pt\relax}
\providecommand{\BIBentryALTinterwordstretchfactor}{4}
\providecommand{\BIBentryALTinterwordspacing}{\spaceskip=\fontdimen2\font plus
\BIBentryALTinterwordstretchfactor\fontdimen3\font minus
  \fontdimen4\font\relax}
\providecommand{\BIBforeignlanguage}[2]{{%
\expandafter\ifx\csname l@#1\endcsname\relax
\typeout{** WARNING: IEEEtran.bst: No hyphenation pattern has been}%
\typeout{** loaded for the language `#1'. Using the pattern for}%
\typeout{** the default language instead.}%
\else
\language=\csname l@#1\endcsname
\fi
#2}}
\providecommand{\BIBdecl}{\relax}
\BIBdecl

\bibitem{ni2023megavul}
\BIBentryALTinterwordspacing
C.~Ni, L.~Shen, X.~Yang, Y.~Zhu, and S.~Wang, ``Megavul: A c/c++ vulnerability
  dataset with comprehensive code representation,'' \emph{arXiv preprint
  arXiv:2406.12415}, 2023. [Online]. Available:
  \url{https://doi.org/10.48550/arXiv.2406.12415}
\BIBentrySTDinterwordspacing

\bibitem{wang2022jtrans}
\BIBentryALTinterwordspacing
H.~Wang, W.~Qu, G.~Katz, W.~Zhu, Z.~Gao, H.~Qiu, J.~Zhuge, and C.~Zhang,
  ``jtrans: Jump-aware transformer for binary code similarity,'' 2022, arXiv
  preprint arXiv:2205.12713. [Online]. Available:
  \url{https://doi.org/10.48550/arXiv.2205.12713}
\BIBentrySTDinterwordspacing

\bibitem{ramos2024large}
\BIBentryALTinterwordspacing
D.~Ramos, C.~Mamede, K.~Jain, P.~Canelas, C.~Gamboa, and C.~Le~Goues, ``Are
  large language models memorizing bug benchmarks?'' \emph{arXiv preprint
  arXiv:2411.13323}, 2024, arXiv preprint arXiv:2411.13323. [Online].
  Available: \url{https://doi.org/10.48550/arXiv.2411.13323}
\BIBentrySTDinterwordspacing

\bibitem{barabosch2023cwechecker}
T.~Barabosch, ``Cwe checker,'' \url{https://github.com/fkie-cad/cwe_checker},
  2023.

\end{thebibliography}

\end{document}